\documentclass[a4paper]{jpconf}
\usepackage{graphicx}
\begin{document}
\title{Neutrinos at extreme energies}

\author{Roberto Aloisio}

\address{Gran Sasso Science Institute (INFN), Viale F. Crispi 7, 67100 L'Aquila, Italy\\
INAF/Osservatorio Astrofisico di Arcetri, largo E. Fermi 5, 50125 Firenze, Italy}

\ead{roberto.aloisio@gssi.infn.it}

\begin{abstract}
We will review the production of neutrinos with PeV energies and above. Discussing two possible sources of this radiation: the propagation of ultra high energy cosmic rays and the decay of super heavy dark matter. The discussion will focus on the theoretical expectations on neutrino fluxes and on the detection capabilities of present and future experiments.
\end{abstract}

\section{Introduction}
The observation of neutrinos with energy as high as $10^{15}$ eV, recently performed by the IceCube detector in Antarctica \cite{Aartsen:2013jdh}, opened up a new window in the observation of the Universe. As always in the history of Science a new observational window will bring new precious informations on known phenomena as well as on unknown new ones. In this paper, following \cite{Aloisio:2015ega} and \cite{Aloisio:2015lva}, we will discuss two important sources of High Energy (HE) neutrinos (with $E\ge 10^{15}$ eV): production due to the propagation of Ultra High Energy Cosmic Rays (UHECR) and to new physics at the inflaton scale. 

UHECR are the most energetic particles observed in nature, with energies up to $10^{20}$ eV. The experimental evidence tells us that UHECR are charged particles, with limits on neutral particles up to $10^{19}$ eV at the level of few percent for photons and well below for neutrinos \cite{Kotera:2011cp}. However, the actual chemical composition of UHECR is still under debate with different experiments claiming contradictory results. 

The Pierre Auger Observatory (Auger) \cite{Aab:2015zoa}, far the largest detector devoted to the observation of UHECR, points toward a mixed composition with light (proton and He) elements dominating the low energy tail of the spectra and a heavier composition at the highest energies, that starts around $E\simeq 5\times 10^{18}$ eV. On the other hand, Telescope Array (TA) \cite{AbuZayyad:2012kk}, even if with 1/10 of the Auger statistics, claims a proton dominated composition at all energies up to the highest observed. 

The production of secondary particles, such as neutrinos or gamma rays, being strongly tied with UHECR chemical composition, can be of paramount importance to solve the alleged contradiction between Auger and TA observations. In this paper we will review secondary neutrino production bracketing the expectations connected with different assumption on UHECR chemical composition. 

UHECR propagation can account for neutrinos of energies as high as $10^{19}$ eV. At the extreme energies, $E\simeq 10^{20}$ eV, the observation of neutrinos can have a dramatic impact on our understanding of the Universe because it could be directly linked with both new physics at the inflation scale and the longstanding problem of Dark Matter (DM). 

The recent claim by BICEP2 of a substantial contribution of tensor modes to the fluctuation pattern of the CMB, even if reconsidered after the combined analysis with Planck and Keck array \cite{Ade:2015tva}, boosted the possible explanation of the DM problem in terms of SHDM, i.e. relic particles created by rapidly varying gravitational fields during inflation (see \cite{Aloisio:2006yi} and reference therein). One of the key expectations of this kind of models is the huge amount of neutrinos (and gamma-rays) produced at energies $\ge10^{20}$ eV \cite{Aloisio:2015lva}. In the present paper we will review recent results that link cosmological observations to the fluxes of UHECR, gamma-rays and neutrinos expected at the highest energies.

\section{Ultra high energy cosmic rays propagation} 
The physics of UHECR propagation is well understood (see \cite{Allard:2011aa} and references therein). During their journey from the source to the observer UHECR experience interactions with the astrophysical photon backgrounds, i.e. the Cosmic Microwave Background (CMB) and the Extragalactic Background Light (EBL). The propagation of UHECR protons\footnote{Hereafter discussing freely propagating UHE nucleons we will always refer only to protons because the decay time of neutrons is much shorter than all other time scales involved.} is affected almost only by the CMB radiation field and the processes that influence the propagation are: (i) pair production and (ii) photo-pion production. On the other hand, the propagation of heavier nuclei is affected also by the EBL and the interaction processes relevant are: (i) pair production and (ii) photo-disintegration. 

The principal source of HE neutrinos is certainly the process of photo-pion production. A nucleon ($N$), whether free or bounded in a nucleus, with Lorentz factor $\Gamma \ge 10^{10}$ interacting with the CMB photons gives rise to the photo-pion production process: $N+\gamma \to N + \pi^0$ and  $N+\gamma \to N + \pi^{\pm}$. At lower energies $\Gamma< 10^{10}$, even if with a lower probability, the same processes can occur on the EBL field. 

\begin{figure}
\begin{center}
\includegraphics[width=0.49\textwidth]{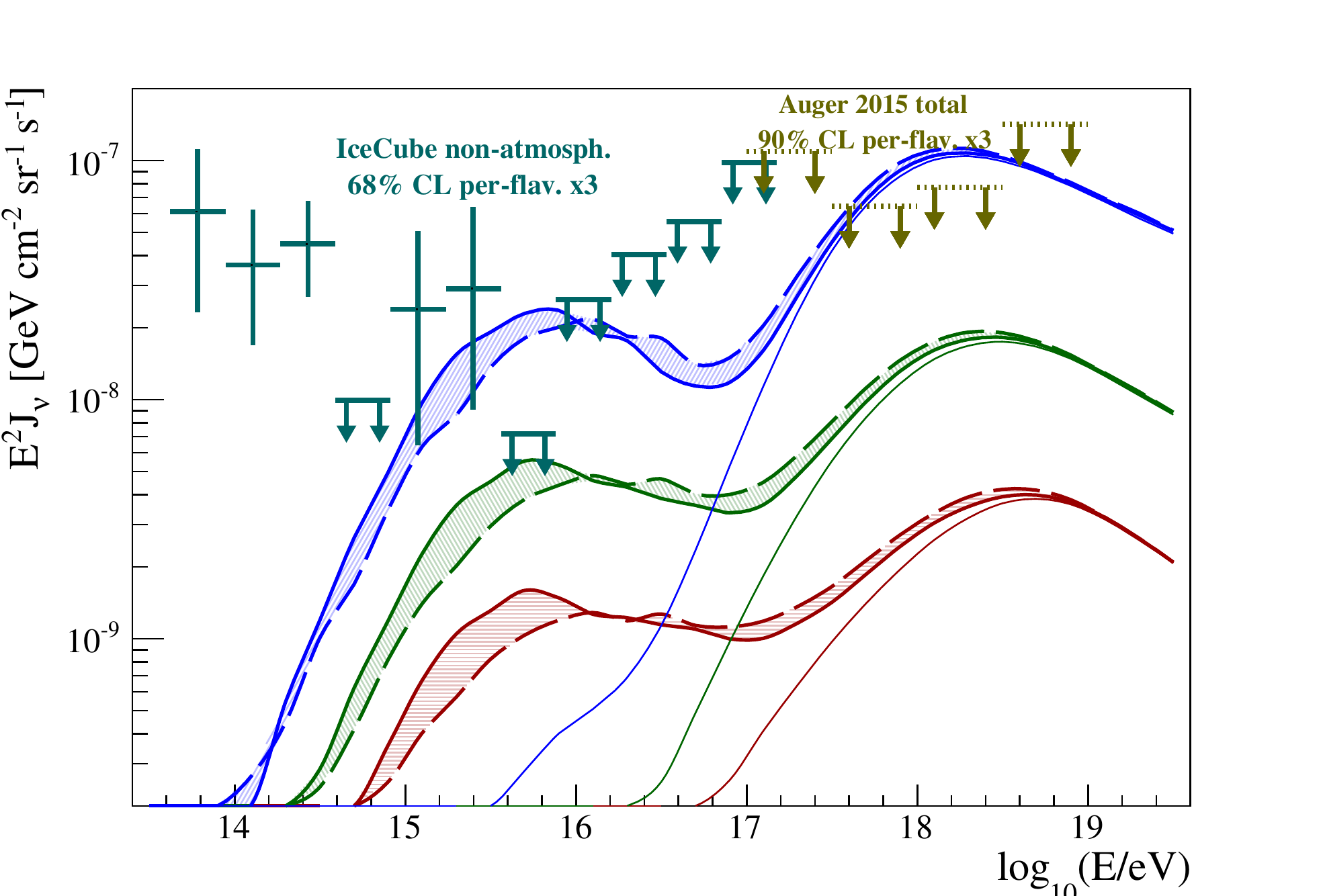}
\includegraphics[width=0.49\textwidth]{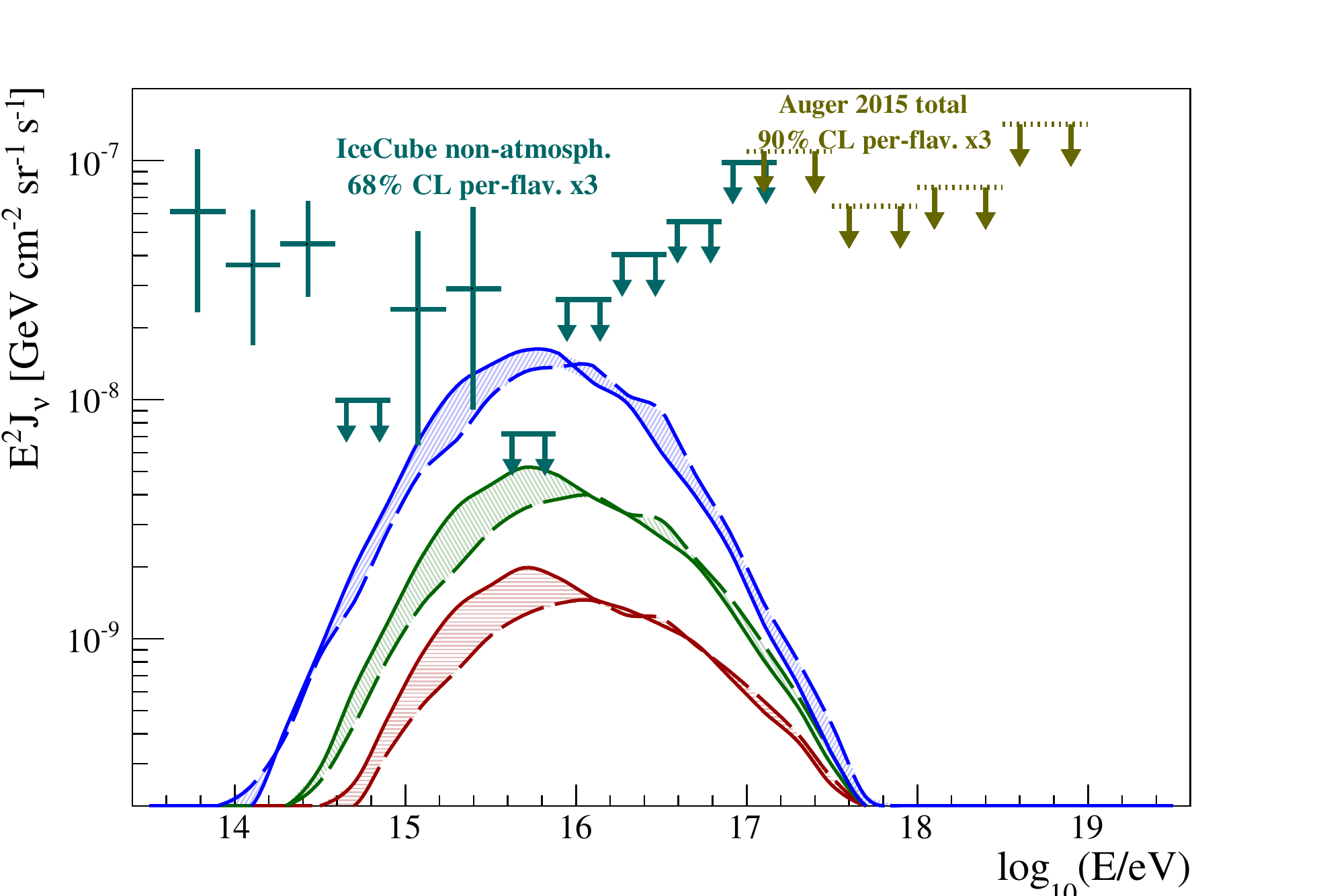}
\caption{\label{fig1} Fluxes of HE neutrinos produced by the propagation of UHECR in the two cases of UHECR's composition: dip model, i.e. pure protons, on the left panel, and mixed composition model, right panel. Experimental data are those of IceCube and Auger, as labeled. The three different bands correspond to three different choices for the cosmological evolution of UHECR sources: red (no evolution) green (SFR evolution) and blue (AGN evolution). Each band shows the uncertainty on the EBL cosmological evolution, solid lines in left panel show neutrino fluxes computed taking into account only the CMB background.}
\end{center}
\end{figure}

The photo-pion production process holds also for nucleons bound within UHE nuclei, being the interacting nucleon ejected from the parent nucleus, but this process is subdominant with respect to nucleus photo-disintegration except at extremely high energies \cite{Allard:2011aa,Aloisio:2008pp,Aloisio:2010he}.

The production of neutrinos comes from the decay of unstable particles (pions, free neutrons and unstable nuclei) produced by the propagation of UHECRs through photo-pion production and photo-disintegration. In most cases the decay length of such particles is much shorter than all other relevant length scales, so these particles decay very soon giving rise to secondary neutrinos.

There are two processes by which neutrinos can be produced in the propagation of UHECRs:
\begin{itemize}
 \item the decay of charged pions produced by photo-pion production, $\pi^{\pm}\to \mu^{\pm} + \nu_\mu(\bar{\nu}_\mu)$, and the subsequent muon decay $\mu^{\pm}\to e^{\pm}+\bar{\nu}_\mu(\nu_\mu)+\nu_e(\bar{\nu}_e)$; 
 \item the beta decay of neutrons and nuclei produced by photo-disintegration: $n \to p + e^{-} + \bar{\nu}_e$, $(A,Z) \to (A,Z-1) + e^{+} + \nu_e$, or $(A,Z) \to (A,Z+1) + e^{-} + \bar{\nu}_e$.
\end{itemize}
These processes produce neutrinos in different energy ranges: in the former the energy of each neutrino is around a few percent of that of the parent nucleon, whereas in the latter it is less than one part per thousand (in the case of neutron decay, larger for certain unstable nuclei). This means that in the interactions with CMB photons, which have a threshold around $\Gamma\ge 10^{10}$, neutrinos are produced with energies of the order of $10^{18}$~eV and $10^{16}$~eV respectively. Interactions with EBL photons contribute with a much lower probability than CMB photons, affecting a small fraction of the propagating protons and nuclei. Neutrinos produced through interactions with EBL, characterised by lower thresholds, have energies of the order of $10^{15}$~eV in the case of photo-pion production and $10^{14}$~eV in the case of neutron decay.   

In figure \ref{fig1} we plot the fluxes of HE neutrinos expected in the case of two different assumptions on the chemical composition of UHECR. In the left panel we plot the case in which UHECR are composed only by protons, the so-called dip model that gives a compelling explanation of the TA data. While in right panel we plot the case of UHECR with a mixed composition, low energies dominated by light elements (p and He) with an heavier composition starting at $E\simeq 5\times 10^{18}$ eV, using the model proposed in \cite{Aloisio:2015ega,Aloisio:2013hya} that gives a good fit of the Auger observations (both spectrum and chemical composition). The three fluxes plotted in figure \ref{fig1} are computed, as detailed in \cite{Aloisio:2015ega}, taking different choices for the cosmological evolution of sources: red (no evolution) green (evolution of the Star Formation Rate) and blue (evolution of Active Galactic Nuclei). Each band shows the uncertainty on the EBL cosmological evolution, solid lines in left panel show neutrino fluxes computed taking into account only the CMB background. From figure \ref{fig1} it is evident the great potential of HE neutrino observations in constraining both UHECR chemical composition and the cosmological evolution of sources. 

\section{Super heavy dark matter}
One of the most fundamental and longstanding problems in modern physics is certainly the presence of a yet-not-observed form of matter whose presence is detected only through its gravitational interaction: the so-called Dark Matter (DM) \cite{Bertone:2004pz}. The leading paradigm to explain DM observations is based on the Weakly Interactive Massive Particle (WIMP) hypothesys, particles of mass of the order of $10^{2}\div 10^{3}$ GeV expected in the context of the ÒnaturalnessÓ argument for electroweak physics \cite{Bertone:2004pz}. 

Searches for WIMP particles are ongoing through three different routes: direct detection, indirect detection, and accelerator searches \cite{Bertone:2004pz}. None of these efforts have discovered a clear WIMP candidate so far. In addition, no evidence for new physics has been observed at the Large Hadron Collider (LHC). Although not yet conclusive, the lack of evidence for WIMPs may imply a different solution for the DM problem outside of the WIMP paradigm.

Following \cite{Aloisio:2015lva}, here we will reconsider the scenario based on particle production due to time varying gravitational fields: the so-called SHDM scenario (see \cite{Aloisio:2015lva,Aloisio:2006yi,Chung:2004nh} and references therein). On general grounds, in the framework of inflationary cosmologies, it was shown that particle creation is a common phenomenon, not tied to any specific cosmological scenario, that can play a crucial role in the solution to the DM problem as SHDM (labeled by X) can have $\Omega_X(t_0)\le 1$ (see \cite{Aloisio:2015lva,Aloisio:2006yi,Chung:2004nh} and references therein). This conclusion can be drawn under three general hypotheses: (i) SHDM in the early Universe never reaches LTE; (ii) SHDM particles have mass of the order of the inflaton mass, $M_\phi$; and (iii) SHDM particles are long-living particles with a lifetime exceeding the age of the Universe, $\tau_X\gg t_0$. These three hypothesis can be tested experimentally through cosmological and UHECR observations. 

\begin{figure}
\begin{center}
\includegraphics[width=0.49\textwidth]{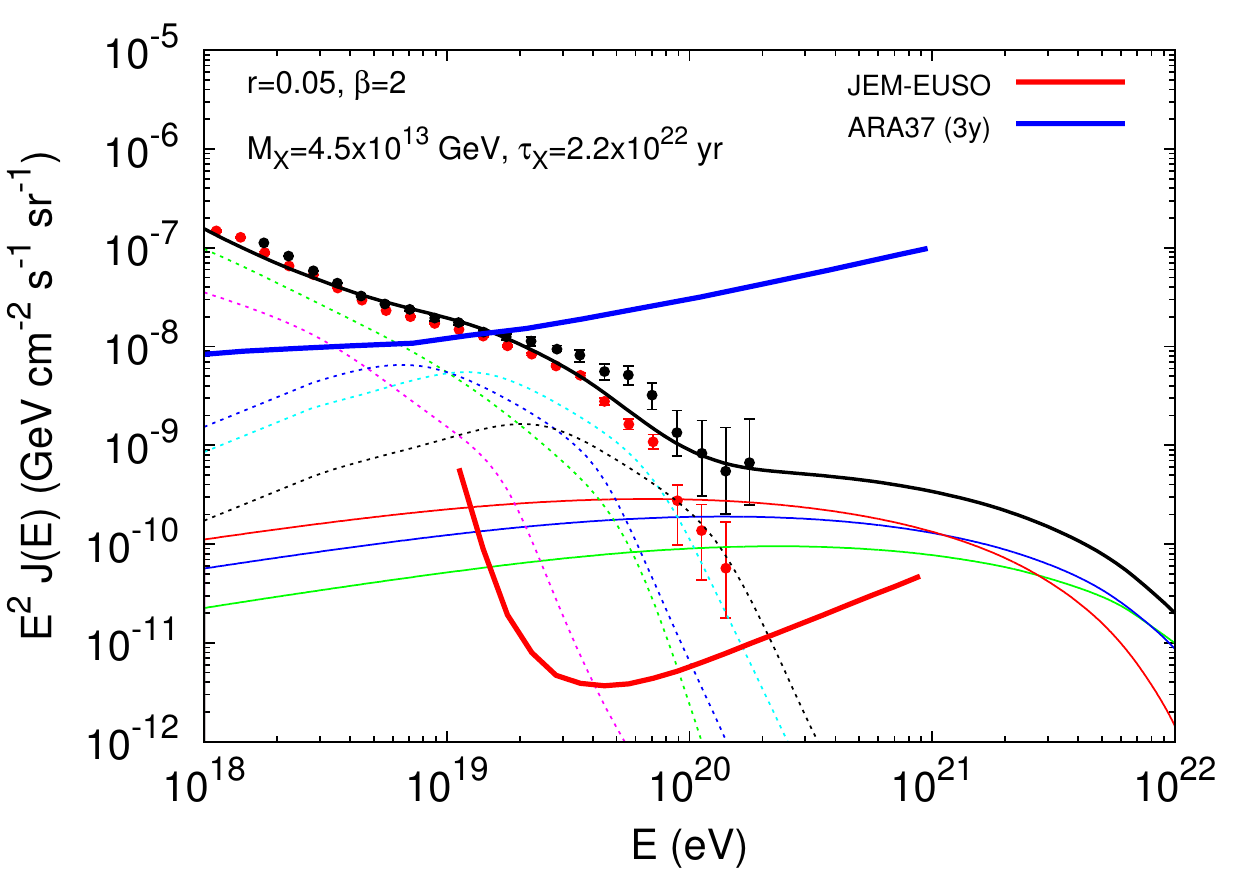}
\includegraphics[width=0.49\textwidth]{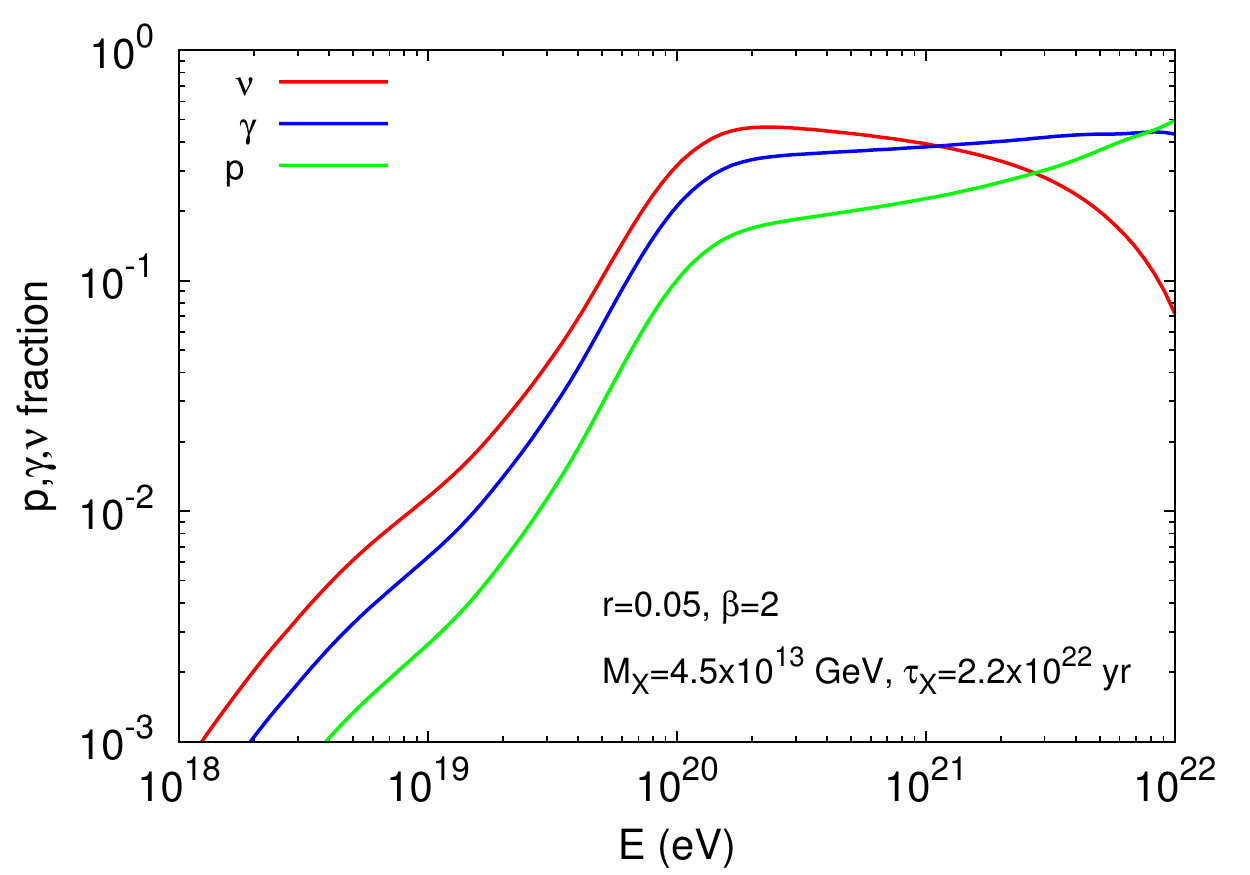}
\caption{\label{fig2} [Left panel] Theoretical fluxes of UHECR from SHDM decay obtained in the case of $r=0.05$ and $\beta=2$, together with UHECR fluxes expected in the framework of the mixed composition model of \cite{Aloisio:2015ega,Aloisio:2013hya}. Also shown: Auger data (red points), TA data (black points) and the sensitivity of the future JEM-EUSO space mission (thick red solid line) and, for UHE neutrinos, the upcoming ARA observatory (thick blue solid line). [Right panel] Ratio over the total UHECR flux of protons, photons and neutrinos by SHDM decay, with the same choice of parameters of the left panel figure. }
\end{center}
\end{figure}

One of the most striking consequences of the SHDM formation is the pollution of the CMB fluctuations by tensor modes \cite{Chung:2004nh}. Therefore the observations performed by BICEP2, analysed together with those of Planck and Keck array \cite{Ade:2015tva}, looking for tensor modes in the CMB fluctuations could boost the SHDM solution. The combined analysis of BICEP2 and Planck showed a ratio of tensor to scalar modes in the CMB background with an upper limit $r\le 0.12$, at $95\%$ confidence level, and a likelihood curve that peaks at $r=0.05$ but disfavours zero with a scarce statistical significance \cite{Ade:2015tva}.  In the near future several different detectors, both ground-based and sub-orbital, will take data enabling a more precise determination of the CMB primordial tensor modes.

Assuming a non-negligible value of the tensor to scalar ratio $r$ in the CMB fluctuations, one gets the scale of the inflaton mass $M_{\phi}$ and, requiring that the SHDM density today corresponds to the observed DM density $\Omega_X=\Omega_{DM}$, one obtains the scale of the SHDM mass $M_X$ \cite{Aloisio:2015lva}. The actual value of $M_X$ also depends on the assumption of the inflaton potential that can be modelled as $V(\phi)=\phi^\beta M_\phi^{4-\beta}/\beta$ with $\beta=2/3,1,4/3,2$ and on the reheating temperature that can be fixed at $10^{9}$ GeV \cite{Aloisio:2015lva}. 

Together with the mass $M_X$ the second parameter that defines the physics of SHDM is its lifetime $\tau_X$. On very general grounds \cite{Aloisio:2006yi}, we can assume that SHDM decay gives rise to a quark anti-quark pair with subsequent parton cascades that, hadronizing, produce Standard Model (SM) particles. The basic signatures of these kind of decays are three: (i) SHDM (as any other DM particle) cluster gravitationally and accumulate in the halo of our Galaxy with an average density $\rho_X^{halo}\simeq 0.3$ GeV/$cm^3$ \cite{Bertone:2004pz}; (ii) in the hadronic cascades the most abundant particles produced are pions, therefore UHE neutrinos and gamma-rays are the most abundant particles expected on Earth; (iii) the non-central position of the Sun in the galactic halo results in an anisotropic flux of the decay products \cite{Aloisio:2007bh}.

In figure \ref{fig2}, left panel, we plot the fluxes of UHECR as in the mixed composition model (dotted lines) discussed in the previous section together with neutrino, photon and proton fluxes coming from SHDM decay (respectively red, blu and green solid lines). The value of $M_X$ chosen corresponds to $r=0.05$ and $\beta=2$, while the SHDM lifetime $\tau_X$ is fixed so to respect the lower limit on gamma-rays fixed by Auger at $E\simeq 10^{19}$ eV. In order to show the detection capabilities of future UHECR or neutrino detectors we have plotted also the expected sensitivities of the JEM-EUSO space mission \cite{Allison:2011wk} (thick red solid line) and, for UHE neutrinos only, the upcoming ARA observatory \cite{Ebisuzaki:2014wka} (thick blue solid line). In right panel of figure \ref{fig2} we show the most important experimental signature of SHDM decay characterised by large fraction of neutrinos and gamma-rays ($\nu/N\simeq 3\div 4$ and $\gamma/N\simeq 2\div 3$). 

From figure \ref{fig2} follows that future observatories of UHECRs and neutrinos should be able to discover SHDM or constrain its lifetimes. The JEM-EUSO detector seems particularly suited for these kind of studies as it achieves about an order of magnitude higher exposure at $10^{20}$ eV respect to Auger.

\section{Conclusions}
We can conclude by stating that the observation of HE neutrinos is an important and solid field of research that could soon bring important new discoveries in high energy astrophysics, cosmology and particle physics. Finally, the present short paper, report of what was presented at the TAUP conference 2015, can also be regarded as a "teaser" to stimulate the interest of the reader in the results obtained in \cite{Aloisio:2015ega} and \cite{Aloisio:2015lva}.

\section*{Aknowledgements}
I'm grateful to all colleagues with whom the results presented here were obtained: D. Boncioli, A. di Matteo, A. Grillo, S. Matarrese, A. Olinto, S. Petrera and F. Salamida. 

\section*{References}
\bibliography{UHECR}
\end{document}